\documentclass[aps,prl,twocolumn,showpacs,preprintnumbers,amsmath,amssymb,superscriptaddress]{revtex4}
\usepackage{graphicx}
\usepackage{tabularx}
\usepackage{color}
\usepackage{xspace}

\begin{document}
\preprint{PREPRINT (\today)}

%
%
\title{Magneto-Electric Coupling in Single Crystal Cu$_2$OSeO$_3$ Studied by a Novel Electron Spin Resonance Technique}
\author{A.~Maisuradze}\email{alexander.m@physik.uzh.ch}
\affiliation{Physik-Institut der Universit\"{a}t Z\"{u}rich, Winterthurerstrasse 190, CH-8057 Z\"{u}rich, Switzerland}
\affiliation{Laboratory for Muon Spin Spectroscopy, Paul Scherrer Institut, CH-5232 Villigen PSI, Switzerland}
\author{A.~Shengelaya}
\affiliation{Department of Physics, Tbilisi State University, Chavchavadze av. 3, GE-0128 Tbilisi, Georgia}
\author{H.~Berger}
\affiliation{Institute of Condensed Matter Physics, \'{E}cole Polytechnique F\'{e}d\'{e}rale de Lausanne (EPFL),
CH-1015 Lausanne, Switzerland}
\author{D. M. Djoki\'c}
\affiliation{Institute of Condensed Matter Physics, \'{E}cole Polytechnique F\'{e}d\'{e}rale de Lausanne (EPFL),
CH-1015 Lausanne, Switzerland}
\author{H.~Keller}
\affiliation{Physik-Institut der Universit\"{a}t Z\"{u}rich, Winterthurerstrasse 190, CH-8057 Z\"{u}rich, Switzerland}

\begin{abstract}
The magneto-electric (ME) coupling on spin-wave resonances in single-crystal Cu$_2$OSeO$_3$ was studied
by a novel  technique using electron spin resonance combined with electric field modulation.
An external electric field ${\bf E}$ induces a magnetic field component $\mu_0 H^i = \gamma E$
along the applied magnetic field ${\bf H}$ with $\gamma=0.7(1)~ \mu$T/(V/mm) at 10 K.
The ME coupling strength $\gamma$ is found to be temperature dependent and highly anisotropic.
$\gamma(T)$ nearly follows that of the spin susceptibility $J^M(T)$
and rapidly decreases above the Curie temperature $T_{\rm c}$. The ratio
$\gamma/J^M$ monotonically decreases with increasing temperature without an anomaly
at $T_{\rm c}$.
\end{abstract}


\maketitle

Magneto-electric (ME) materials exhibiting coupled and microscopically
coexisting magnetic (${\bf M}$) and electric (${\bf P}$) polarizations have attracted considerable
interest in recent years \cite{Hill00, Eerenstein, Spaldin05, Fiebig05}.
This coupling allows one to influence the magnetic state of a ME material via an external
electric field, thus opening a broad range of possible technical applications of such
materials \cite{Eerenstein, Bibes08}. Moreover, it is very interesting to investigate the
microscopic mechanism of ME coupling, since ${\bf P}$ and ${\bf M}$
tend to exclude each other \cite{Hill00}. In order to detect the ME effect, sensitive
and reliable experimental techniques are required, since this coupling is generally quite small.
Usually, for the determination of the ME
coupling either the dielectric properties of ME materials are measured as a function of magnetic
field or the magnetization is studied as a function of an applied electric field \cite{Eerenstein}.

Cu$_2$OSeO$_3$ is a paraelectric ferrimagnetic material with a Curie
temperature of $T_{\rm c}\simeq 57$ K \cite{Bos08,Effenberger86,Maisuradze11}.
The ME effect in Cu$_2$OSeO$_3$ was first observed by magneto-capacitance
experiments \cite{Bos08}. Later on, a small abrupt change of
the dielectric constant below $T_{\rm c}$ was reported by infrared
reflection and transmission studies \cite{Miller10,Gnezdilov10}.
Recent $\mu$SR investigations showed a rather small change of the internal magnetic
field by applying an electric field \cite{Maisuradze11}. X-ray diffraction \cite{Bos08}
and nuclear magnetic resonance \cite{Belesi10} studies
revealed no evidence of any lattice anomaly below $T_{\rm c}$, suggesting that lattice
degrees of freedom are not directly involved in the ME effect. Moreover, a metastable
magnetic transition with enhanced magneto-capacitance was observed \cite{Bos08} and later on
was also investigated under hydrostatic pressure \cite{Huang11}. Very recently,
ME skyrmions  were observed in Cu$_2$OSeO$_3$ by means of Lorentz transmission
electron microscopy \cite{Seki12} and small angle neutron
scattering \cite{Adams12}.

Here we report a study of the ME coupling in single crystal of Cu$_2$OSeO$_3$.
For this investigation a novel microscopic method for the direct determination
of the ME effect based on the standard FMR/EPR technique combined with electric field
modulation was developed.
As a result, to our knowledge for the first time spin-wave
resonance (SWR) excitations \cite{Kittel58} were detected via ME coupling.
The linear ME coupling strength $\gamma$ was determined quantitatively in Cu$_2$OSeO$_3$.
In particular, the temperature and angular dependence of $\gamma$ and the SWR
excitations were investigated.
The temperature dependence of the ME coupling was found to follow nearly that of
the spin susceptibility without a sudden change across T$_{\rm c}$.
By comparing the results of ME Cu$_2$OSeO$_3$ with those of standard DPPH
(C$_{18}$H$_{12}$N$_5$O$_6$) we further demonstrate that this novel microscopic
method is a very sensitive and powerful tool to investigate the ME effect and
to search for new ME materials.

\begin{figure}[tb]
\includegraphics[trim=1.5cm 1.9cm 3.1cm 0.9cm, clip=true,width=0.25\linewidth]{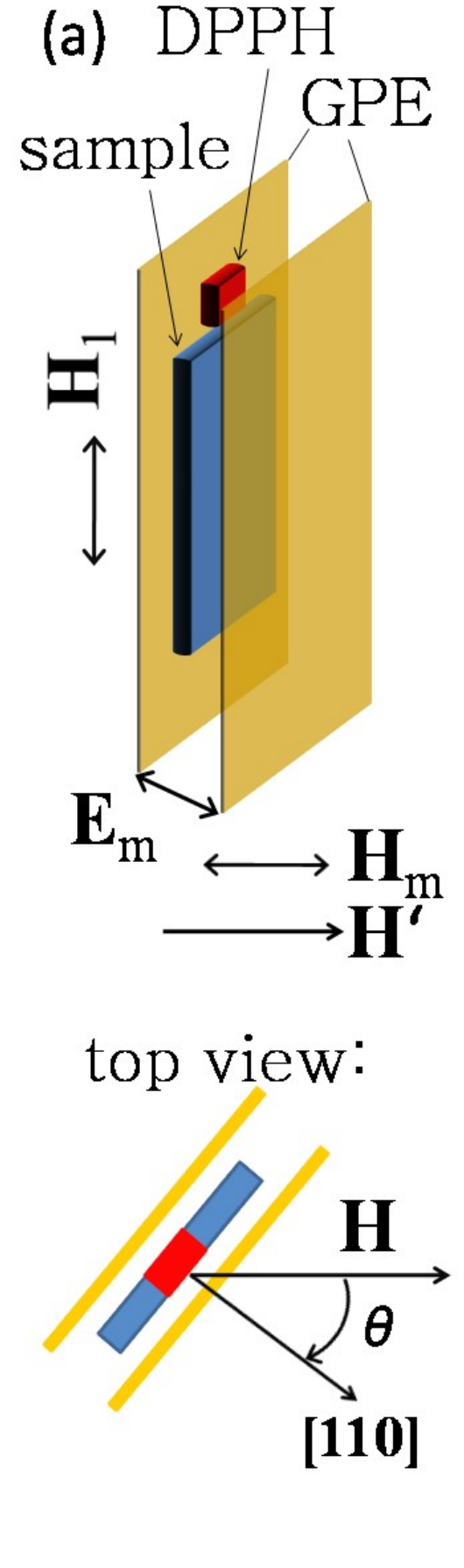}
\includegraphics[width=0.72\linewidth,trim=0.4cm 0.4cm 0.8cm 0.5cm, clip=true]{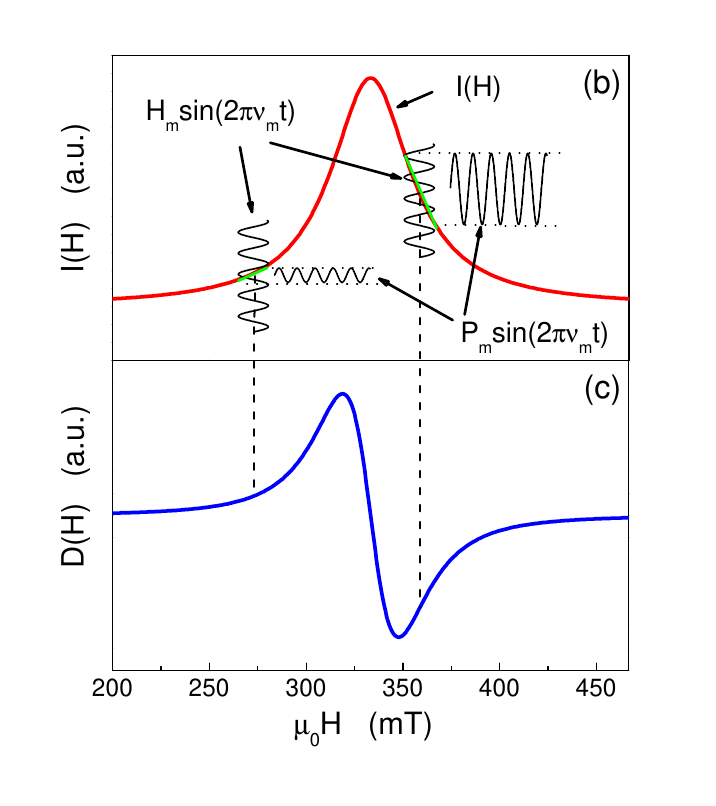}
\caption{ \label{fig:EPRdetection} (Color online) (a) Schematic view of the sample and the magnetic/electric field geometry. The ME sample and the marker sample DPPH are sandwiched between two gold plate electrodes  (GPE).  ${\bf H'}$: static external magnetic field,
${\bf H_{\rm m}}$: magnetic modulation field, ${\bf E_{\rm m}}$: electric modulation field, ${\bf H_{1}}$: microwave field,
${\bf H} = {\bf H'} + {\bf H_{\rm m}}$: total external magnetic field.
(b) Basic principle of EPR signal detection. Red curve represents the EPR absorption line  $I(H)$.
The modulation magnetic field $H_m\sin(2\pi\nu_m t)$ and the resulting modulated microwave
absorption power $P_m\sin(2\pi\nu_m t)$ are also illustrated.
(c) First derivative $D(H)$ signal of the EPR absorption line $I(H)$ after lock-in detection.}
\end{figure}

High quality single crystals of Cu$_2$OSeO$_3$ were prepared using a procedure
described elsewhere \cite{Belesi10}. The crystal structure is cubic with symmetry
(P$2_13$) \cite{Bos08,Effenberger86}.  Several thin single-crystal samples of approximate
dimensions of  $\sim 1\times 1\times d$~mm with thickness $d\leq 0.1$~mm were studied.
The $[110]$ direction of the crystal is oriented perpedicular to the planes of the thin
samples (see Fig. \ref{fig:EPRdetection}a).
The FMR and EPR measurements were performed with a standard
X-band (9.6 GHz) {\em BRUKER EMX} spectrometer. In order to detect the ME
effect a capacitor-like structure consisting of two thin ($< 10$~$\mu$m) isolated gold electrodes
separated by $\simeq 0.3$ mm was used (see Fig. \ref{fig:EPRdetection}a).
The sample and the DPPH marker \cite{AbragamAndBleney70, Poole97}
were placed between the two electrodes. The electrodes were connected to an ac
voltage source of amplitude $V_m = 17$ V, synchronized with the frequency of 100~kHz of the
magnetic field modulation generator of the spectrometer  \cite{AbragamAndBleney70, Poole97}.
Two kind of resonance experiments
were performed: (1) EPR/FMR with standard magnetic field modulation (MFM) and
(2) EPR/FMR with electric field modulation (EFM) at 100 kHz.
The $[110]$ axis of the crystal was perpendicular to the microwave field ${\bf H_1}$.
For the angular dependent measurements the sample was rotated with respect to the applied
field ${\bf H}$ (see Fig. \ref{fig:EPRdetection}a).
\begin{figure}[tb]
\includegraphics[width=0.89\linewidth,trim=0.4cm 0.4cm 0.8cm 0.8cm, clip=true]{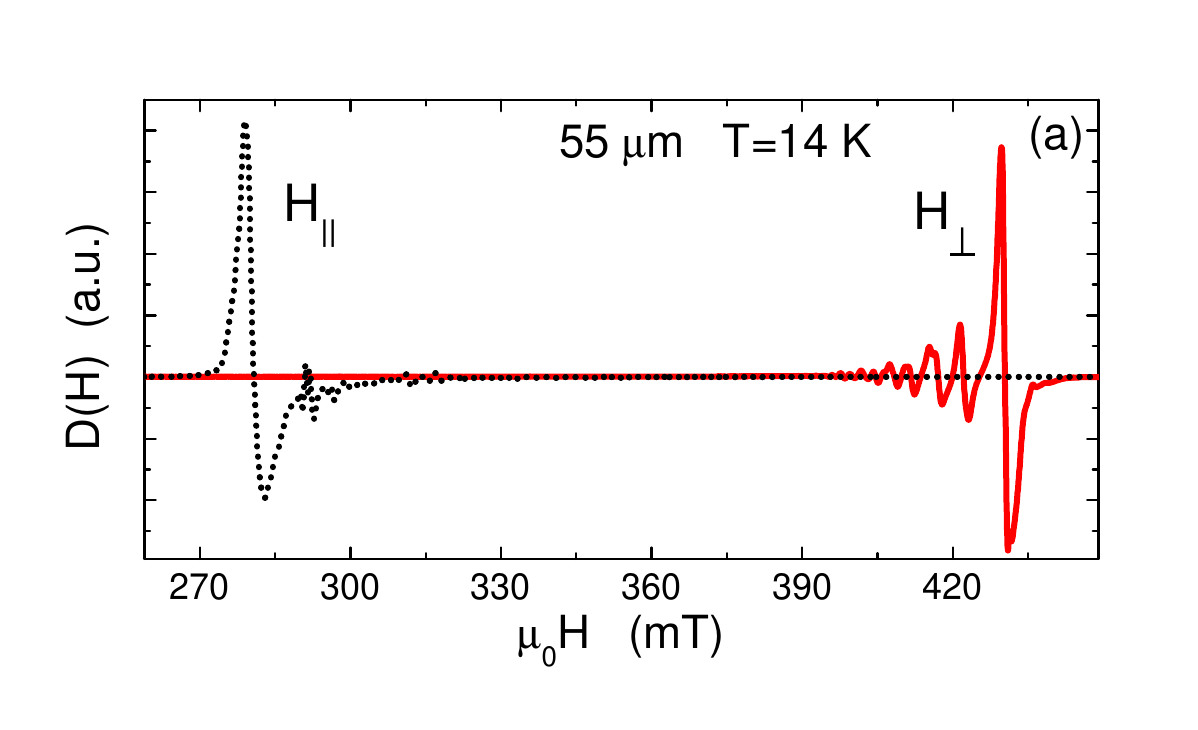}
\includegraphics[width=0.89\linewidth,trim=0.4cm 0.4cm 0.8cm 0.8cm, clip=true]{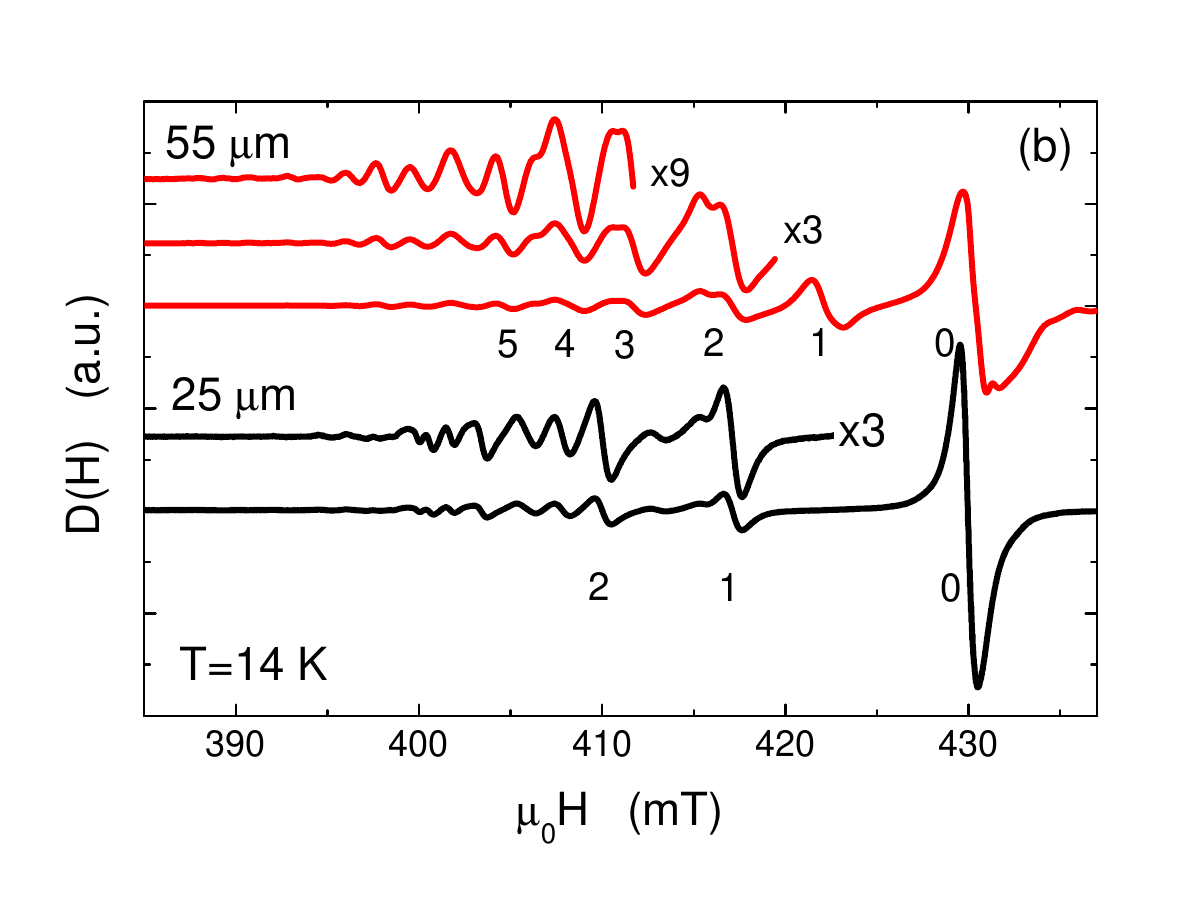}
\caption{ (Color online) (a) FMR in the 55 $\mu$m thick single-crystal sample of Cu$_2$OSeO$_3$ at $T = 14$~K
for $H_{||}$ and $H_{\perp}$ using the conventional MFM technique. (b) SWR in the 25 and 55 $\mu$m thick single-crystal samples of Cu$_2$OSeO$_3$ at $T = 14$~K for $H_{\perp}$. The indices 0, 1, 2, ... indicate the order of the SWR mode. }
\label{fig:SignalAt14K}
\end{figure}

The EPR/FMR technique is based on the resonance absorption of microwave energy
by a Zeeman-split spin system \cite{AbragamAndBleney70, Poole97}.
The Zeeman splitting of the spin system is achieved by sweeping an applied magnetic field $H$.
In the simplest case of an effective spin $S=1/2$ system (as for the present case of Cu$^{2+}$) the
double degenerate ground state is split into two levels by the Zeeman energy $E_Z = g \mu_B H$.
When $E_Z = h \nu$, where $\nu = 9.6$ GHz is the frequency of the microwave $H_1$ field,
resonance absorption takes place (see Fig. \ref{fig:EPRdetection}b).
In order to increase the sensitivity, the applied magnetic
field $H$ is modulated: $H = H' + H_m\sin(2\pi\nu_m t)$, where $H'$ is the static applied magnetic field,
$H_m$ is the modulation amplitude, and $\nu_m$ is the modulation frequency (typically $\nu_m=100$ kHz).
During signal detection $H'$ is swept slowly. As a result, the detected microwave absorption power
$P(t) = P_{m}\sin(2\pi\nu_m t)$ is also modulated with the frequency $\nu_m$.
The amplitude $P_{m}$ is proportional to the slope
$D(H)$ of the absorption signal $I(H)$. Further amplification and lock-in detection of
$P(t)$ results in the EPR derivative signal $D(H)$ as illustrated
in Fig. \ref{fig:EPRdetection}c \cite{AbragamAndBleney70, Poole97}.

Ferromagnetic resonance studies were performed previously on composite ME structures
involving piezoelectric and magnetostrictive compounds \cite{Chen11, Weiler11,Shastry04}.
In these experiments a static external field $E_{\rm st}$ was used to detect the ME coupling strength.
By applying $E_{\rm st}=1$ kV/mm in the present experiments a shift of the resonance fields of the
order of $\simeq0.5$ mT was also detected for Cu$_2$OSeO$_3$, indicating an additional
magnetization induced by $E_{\rm st}$ (see supplemental materials). However, in order to
increase the sensitivity of signal detection and to avoid artifacts related to
hysteresis effects of the magnet core it is advantageous to apply
a periodic voltage to detect small changes in the spectra.
This technique was previously applied to investigate the electric field effect
on the non-Kramers ion Pr$^{3+}$ in LaMgN$_2$\cite{Whysling76}.
The main idea of the present experiment is to use EFM to observe EPR/FMR
signals in Cu$_2$OSeO$_3$ instead of the usual MFM technique.
In a spin system without ME effect (e.g. DPPH) no modulated signal
$P(t) = P_{m}\sin(2\pi\nu_m t)$ will occur. However, if the ME effect is present in the sample,
modulation by an electric field $E_m \sin(2\pi \nu_m t)$ leads to a modulation of the
magnetization $M(t)$ and therefore to a modulation of the magnetic field in the sample
$B(t) = \mu_0[H+M(t)]$. In this case the EPR/FMR signal
which is proportional to the ME coupling may in principle be detected.

\begin{figure}[tb]
\includegraphics[width=0.89\linewidth,trim=0.4cm 0.4cm 0.8cm 0.8cm, clip=true]{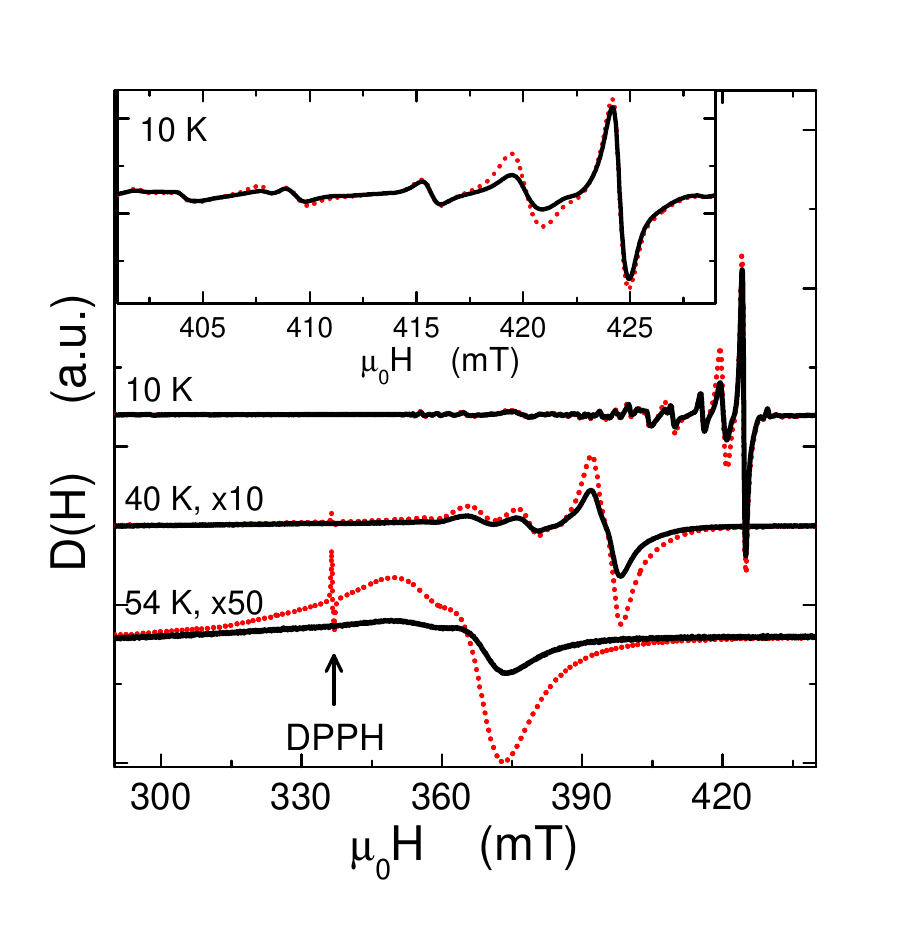}
\caption{ (Color online)
Temperature dependence of SWR signals of single-crystal Cu$_2$OSeO$_3$ detected using
the MFM technique (dotted line) and the EFM technique (solid line). The sharp peak visible
at 340~mT and 54~K is the signal of the marker sample DPPH
which is present only in the case of MFM. The inset shows the expanded spectra at 10 K around 415~mT.
For better comparison all the EFM signals are multiplied by a factor of 2.}
\label{fig:SignalTdep}
\end{figure}

First we describe the FMR/EPR signals obtained
in Cu$_2$OSeO$_3$ using the conventional MFM technique. For a polycrystalline or
arbitrarily shaped single crystal a very complex signal is observed as reported previously (see Ref. \cite{Kobets10}).
We found that the signal is substantially simpler for a thin single crystal with a
nearly constant effective demagnetization factor \cite{Zheng96}.
Figure \ref{fig:SignalAt14K}a shows the FMR signal of a thin single-crystal sample of Cu$_2$OSeO$_3$
(thickness $d=55$~$\mu$m) at 14 K with the applied magnetic field $H$ parallel ($H_{||},\theta = 90^{\circ}$) and
perpendicular ($H_{\perp}, \theta = 0^{\circ}$) to the plane of the sample (see Fig. \ref{fig:EPRdetection}a).
For $H_{||}$ a slightly skewed single signal
is observed, whereas for $H_{\perp}$ multiple peaks with different
signal intensities are evident. These peaks represent resonances of different
spin-wave (SW) modes. In thin ferromagnetic samples SW modes are expected to occur at
resonance fields $H_n$ \cite{Kittel58, Rappoport04, Liu06}:
\begin{equation}\label{eq:spinWHn}
H_n=  H_0 - S\left(\frac{\pi}{d}\right)^2[(n+1)^2-1]
\end{equation}
Here, $S$ is a parameter related to the spin stiffness \cite{Kittel58, Rappoport04},
$n$ is the order of the spin-wave mode, and $d$ is the thickness of the sample.
With increasing $n$ the resonance field $H_n$ decreases, and
with decreasing $d$ the difference $H_0-H_1$ increases. Qualitatively this behavior agrees
with our observation (see Fig. \ref{fig:SignalAt14K}b). However, there are quantitative
deviations from Eq. (\ref{eq:spinWHn}) as was reported previously for various materials \cite{Rappoport04, Liu06}.
These deviations are often related to stress, magnetic anisotropy, distribution or
variation of magnetization across the sample. As shown in Fig. \ref{fig:SignalAt14K} it was possible
to detect SW modes with order $n >10$ in the present experiments.
However, due to a slight variation of thickness $d$ across the
sample the modes of high order $n$ interfere. Moreover, with increasing temperature
the linewidths of the SWR modes increase and overlap (see Fig. \ref{fig:SignalTdep}).

\begin{figure}[!t]
\includegraphics[width=0.85\linewidth,trim=0.4cm 0.4cm 0.8cm 0.8cm, clip=true]{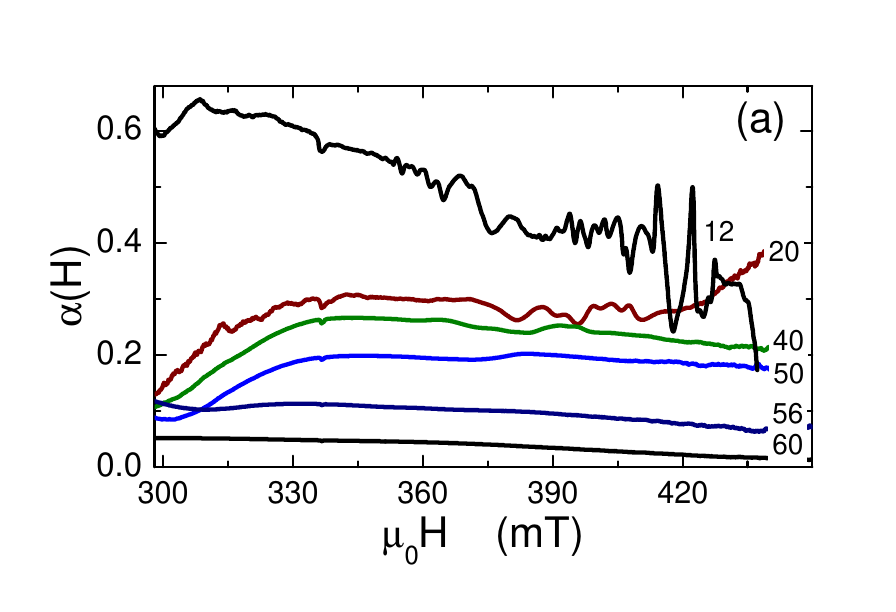} 
\includegraphics[width=0.85\linewidth,trim=0.4cm 0.4cm 0.7cm 0.8cm, clip=true]{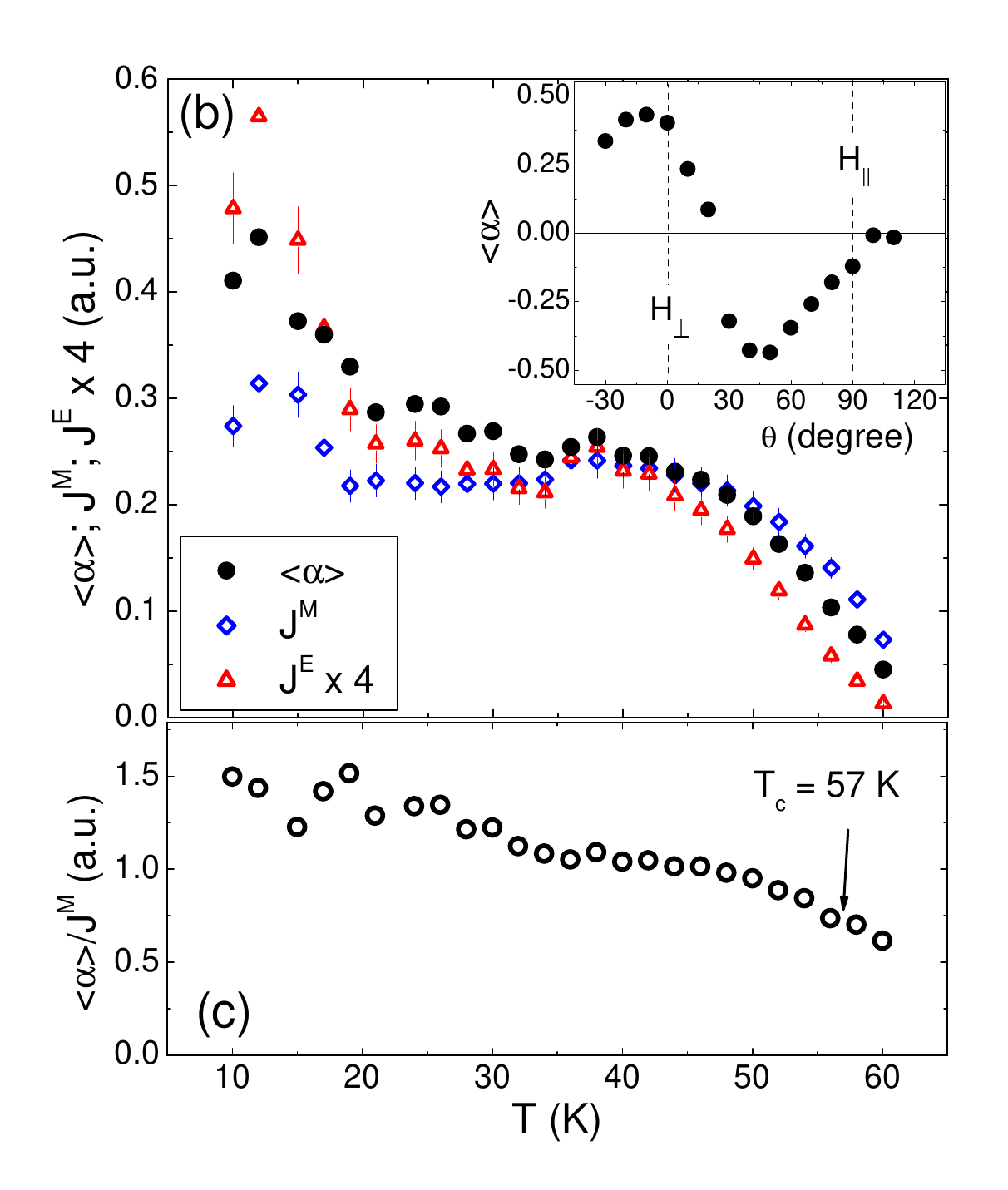}  
\caption{ (Color online) (a) Spectrally resolved ME effect parameter $\alpha(H)$
in single-crystal Cu$_2$OSeO$_3$ at $T = 12$, 20, 40, 50, 56, and 60 K for $H_{\perp}$.
(b) Temperature dependence of the average ME effect parameter $\langle\alpha\rangle$
and the signal intensities $J^E$ and $J^M$ detected by EFM and MFM, respectively.
The inset shows the angular dependence of $\langle\alpha\rangle$ at 14 K.
(c) Temperature dependence of the ratio $\langle \alpha\rangle / J^M$ showing no anomaly at~$T_{\rm c}$.  }
\label{fig:MEresults}
\end{figure}

\begin{figure}[!t]
\includegraphics[width=0.85\linewidth,trim=0.4cm 0.4cm 0.8cm 0.8cm, clip=true]{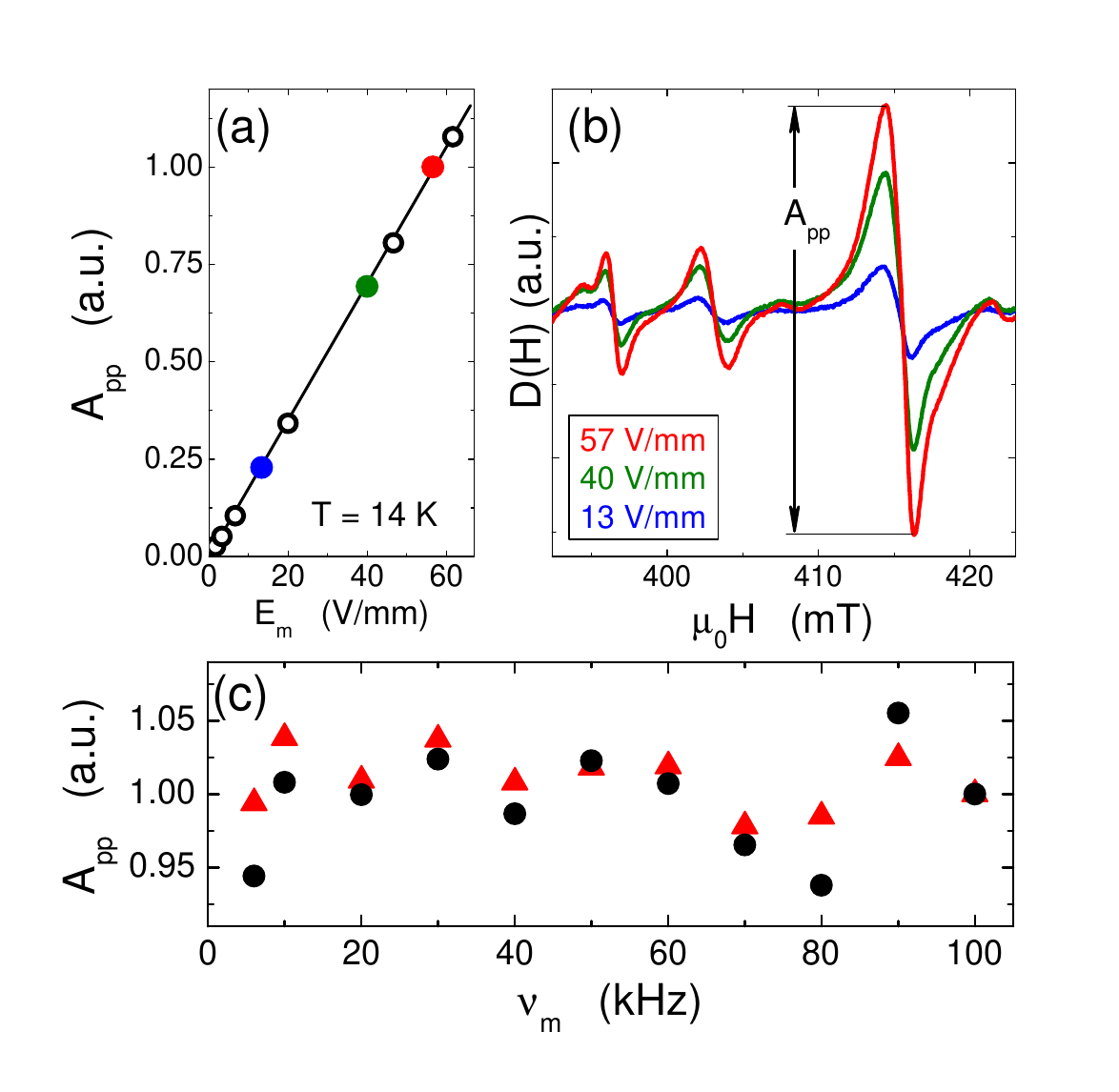} 
\caption{(Color online) (a) Peak-to-peak amplitude $A_{\rm pp}$ of the zero order
SWR mode extracted from $D(H)$ shown in (b) as a function of the EFM amplitude $E_m$ in Cu$_2$OSeO$_3$.
Note that $A_{\rm pp}\propto E_m$ indicating that ME coupling is linear.
(c) $A_{\rm pp}$ as a function of the EFM frequency $\nu_m$ at $T=25$ K ($\bullet$)
and 35 K ($\blacktriangle$).  }
\label{fig:Fig5}
\end{figure}

Next we discuss the ME effect using the EFM method by applying an ac electric field $E_m\sin(2\pi\nu_m t)$.
Figure \ref{fig:SignalTdep} shows some typical FMR spectra of the 55 $\mu$m thick
single-crystal sample at different temperatures detected by this technique. It is evident
that the SWR lines are also observed as for MFM. At 10 K the SWR signals $D^{M}(H)$ and
$D^{E}(H)$ detected by MFM and EFM, respectively,
have approximately the same amplitudes. With increasing temperature, however, the amplitude of
$D^{E}(H)$ is reduced compared to that of $D^{M}(H)$. Above 60 K the intensity of the
$D^{E}(H)$ signal becomes very small.  Note that for the marker sample DPPH no signal is present
in the case of EFM as expected. The absence of a DPPH signal unambiguously demonstrates that ME coupling in
Cu$_2$OSeO$_3$ gives rise to the $D^{E}(H)$ signal (see also supplemental materials).
Thus, the ratio of the signal intensities detected by electric and magnetic modulations
is proportional to strength of the ME effect \cite{Comment1}:
\begin{equation}\label{eq:MEeffect}
\alpha(H) = \frac{I^E(H)}{I^M(H)}\equiv \frac{\int_0^H D^E(h)dh}{\int_0^H D^M(h)dh}.
\end{equation}
This ratio is determined by $\alpha=\mu_0H^i/\mu_0H_m$, where
$\mu_0H^i_m=\gamma E_m$ is the magnetic field induced by the electric field $E_m$ and
$\mu_0H_m=0.1$ mT is the field used in the MFM experiment. Therefore, the ME coupling strength
$\gamma=\alpha C$  with a calibration factor
$C=\mu_0H_m/E_m = 1.76$ $\mu$T/(V/mm) \cite{Eerenstein,Fiebig05}.
It is convenient to introduce
the spectrally averaged value of $\alpha(H)$:
\begin{equation}
\langle\alpha\rangle = \frac{\int I^M(H)\alpha(H)dH}{\int I^M(H)dH}=\frac{\int I^E(H)dH}{\int I^M(H)dH}=\frac{J^E}{J^M}.
\end{equation}
Here,  $J^{\beta}=\int\int D^{\beta}(h)d^2h$, ($\beta = E, M$) are the signal intensities in the
EFM and the MFM experiments, respectively. $J^{M}$ is also a measure of spin susceptibility \cite{AbragamAndBleney70}.
The temperature dependence of $\alpha(H)$ for $H_{\perp}$ is shown in Fig. \ref{fig:MEresults}a.
Above 20 K  $\alpha(H)$ has a minimum at around 300 mT, and is nearly constant above 320 mT.
The averaged ME effect parameter $\langle\alpha\rangle$ as a function of temperature is plotted in
Fig. \ref{fig:MEresults}b, together with the temperature dependencies of  $J^E$ and $J^M$.
The ME effect is most pronounced at low temperatures
and decreases with increasing temperature.
With the above value of $C$ and $\langle \alpha\rangle=0.4$ one obtains $\gamma=0.7(1)$ $\mu$T/(V/mm) at 10 K.
Below 20 K the ME effect
is decreasing slightly faster than $J^M$ with increasing temperature as shown in Fig. \ref{fig:MEresults}b.
Above 60 K the ME effect rapidly decreases, although it is still present
in the paramagnetic phase. The insert of Fig. \ref{fig:MEresults}b shows $\langle\alpha\rangle$ at 14 K
as a function of the angle $\theta$ (see Fig. \ref{fig:EPRdetection}a).
The sign change of $\langle\alpha(\theta)\rangle$ at $\theta \simeq 25^{\circ}$ corresponds to the
change of the direction of the
induced magnetization $M^i$ with respect to ${\bf H}$. The observed $\langle\alpha(\theta)\rangle$
indicates that the ME effect depends not only on the relative orientation of ${\bf E_{\rm m}}$ and
${\bf H}$ (see Fig. \ref{fig:EPRdetection}a), but also on the crystal orientation with respect to these fields.
In Fig. \ref{fig:MEresults}c the temperature dependence of the ratio
$\langle\alpha\rangle /J^M$ is shown. This ratio decreases gradually with increasing temperature
showing no anomaly at $T_{\rm c}=57$ K indicating that ME coupling mechanism is not related to
the onset of long range magnetic order.
The ME coupling is linear as is evident from the linear relation between the  SWR
peak-to-peak amplitude $A_{\rm pp}$ of $D(H)$
and the applied EFM amplitude $E_m$ (see Figs. \ref{fig:Fig5}a and \ref{fig:Fig5}b). In Fig. \ref{fig:Fig5}c
we show $A_{\rm pp}$ as a function of the EFM frequency $\nu_m$.
Note that $A_{\rm pp}$ shows no appreciable frequency dependence, indicating that
$\gamma$ is not related to a mechanical resonance of the sample, as observed in some of
the composite ME materials \cite{Laletsin04}.

It is interesting to compare the magnitude of the ME effect $\gamma=\alpha C$ observed in this work
with previous results \cite{Bos08, Maisuradze11}. For an applied electric field of
$\delta E = 500/3$ (V/mm) a change of the internal magnetic field of
$\mu_0 \delta \overline{H^i_{\mu\rm SR}}= 0.4(4)$ mT was detected by $\mu$SR \cite{Maisuradze11}.
This corresponds to an electric field induced magnetization of
$\mu_0 \delta\overline{ M_{\mu\rm SR}} = \mu_0 \delta \overline{H^i_{\mu\rm SR}}(1-N)^{-1} \simeq 0.6$ mT
(for $N=1/3$ \cite{Comment2}).
For the same electric field and for a mean value of $\overline{\langle\alpha\rangle} \simeq 0.28$
for $T < 50$~K, the average induced magnetization is estimated to be
$\mu_0 \delta \overline{M_{\rm FMR}}= \mu_0 \delta\overline{H^i_{\rm FMR}}(1-N)^{-1} =
\overline{\langle\alpha\rangle} C \delta E(1-N)^{-1} \simeq 0.55$ mT, where $N\simeq 0.85$ was used
corresponding to the actual geometry of the sample \cite{Zheng96}.
The present value of $\mu_0 \delta \overline{M_{\rm FMR}}\simeq 0.55$~mT is in good agreement with the
value of $\simeq 0.6$~mT obtained by $\mu$SR \cite{Maisuradze11}.
The observed temperature dependence of the ME effect differs slightly from that measured by
magneto-capacitance experiments on powder samples \cite{Bos08}, but it is in agreement with that
observed recently for a single crystal sample \cite{Belesi12}.
While the ME effect parameter $\langle\alpha\rangle$ is
strongly reduced above 60 K, the ME effect reported in Ref. \cite{Bos08} is still substantial up to 65~K.

In summary,  the magneto-electric coupling in single crystal of Cu$_2$OSeO$_3$
was studied by means of a novel and highly sensitive magnetic resonance technique.
This method is based on the use of electric field modulation
instead of conventional magnetic field modulation in standard
continuous wave EPR.
Resonance lines of spin wave modes of more than order 10 could be resolved in the FMR spectra. Moreover, spin
wave resonances were observed via the ME coupling by applying an electric field modulation technique.
By combining magnetic and electric field modulation experiments, the temperature and
angular dependence of the linear ME effect in Cu$_2$OSeO$_3$ was investigated for the electric field parallel
to the $[110]$ direction of the crystal.
The ME coupling was found to be $\gamma=0.7(1)$ $\mu$T/(V/mm) at 10 K.
The magnetization induced by the applied electric field is in
good agreement with previous $\mu$SR results \cite{Maisuradze11}.
The temperature dependence of the ratio of ME coupling strength to the spin susceptibility
$\gamma/J^M$ exhibits no anomaly at $T_{\rm c}=57$ K. This indicates that the ME coupling mechanism
is not related to the presence of long range  magnetic order.

We acknowledge support by the Swiss National
Science Foundation, the NCCR Project {\em Materials with Novel
Electronic Properties} (MaNEP), the SCOPES Grant No.
IZ73Z0-128242, and the Georgian National Science Foundation
Grant GNSF/ST08/4-416.
%
%


\begin{thebibliography}{99}

\bibitem{Fiebig05} M. Fiebig,
J. Phys. D: Appl. Phys. {\bf 38} R123 (2005).

\bibitem{Spaldin05} N. A. Spaldin and M. Fiebig,
Science {\bf 309}, 391 (2005).

\bibitem{Eerenstein} W. Eerenstein, N.D. Mathur, and J.F. Scott,
Nature {\bf 442}, 759 (2006).

\bibitem{Hill00} N. A. Hill,
J. Phys. Chem. B {\bf 104}, 6694 (2000).

\bibitem{Bibes08} M. Bibes and A. Barth\'el\'emy,
Nature Materials {\bf 7}, 425 (2008).


\bibitem{Bos08}
J.-W. G. Bos, C. V. Colin, and T. T. M. Palstra,
Phys. Rev. B {\bf 78}, 094416 (2008).

\bibitem{Effenberger86} H. Effenberger and F. Pertlik,
Monatsch. Chem. {\bf 117}, 887 (1986).

\bibitem{Maisuradze11}
A. Maisuradze {et al.},
Phys. Rev. B {\bf 84}, 064433 (2011).

\bibitem{Miller10}
K. H. Miller {\it et al.},
Phys. Rev. B {\bf 82}, 144107 (2010).

\bibitem{Gnezdilov10} V. Gnezdilov {\it et al.},
Fiz.Nizk. Temp. {\bf 36}, 688 (2010).

\bibitem{Belesi10} M. Belesi {\it et al.},
Phys. Rev. B {\bf 82}, 094422 (2010).

\bibitem{Huang11} C.L. Huang {\it et al.},
Phys. Rev. B {\bf 83}, 052402 (2011).

\bibitem{Seki12} S. Seki, X. Z. Yu, S. Ishiwata, and Y. Tokura,
Science {\bf 336}, 198 (2012).

\bibitem{Adams12} T. Adams {\it at al.}, 	arXiv:1204.3597v1 (2012).

\bibitem{Kittel58} C. Kittel,
Phys. Rev. {\bf 110}, 1295 (1958).

\bibitem{AbragamAndBleney70} A. Abragam and B. Bleaney,
{\it Electron Paramagnetic Resonance of Transition Ions} (Clarendon, Oxford, 1970).

\bibitem{Poole97} Ch. P. Poole,
{\it Electron Spin Resonance: A Comprehensive Treatise on Experimental Techniques}
(Dover Publications, 2 ed. 1997).

\bibitem{Chen11}  Y. Chen {\it et al.},
Phys. Rev. B {\bf 83}, 104406 (2011).

\bibitem{Weiler11}  M. Weiler {\it et al.},
Phys. Rev. Lett. {\bf 106}, 117601 (2011).

\bibitem{Shastry04} S. Shastry, {\it et al.},
Phys. Rev. B {\bf 70}, 064416 (2004).

\bibitem{Whysling76} P. Wysling and K.A. M\"uller,
J. Phys. C: Solid State Phys. {\bf 9}, 635 (1976).

\bibitem{Kobets10} M.I. Kobets {\it et al.},
Low. Temp. Phys. {\bf 36}, 176 (2010).

\bibitem{Zheng96}  G. Zheng {\it et al.},
J. Appl. Phys. {\bf 79}, 5742 (1996).

\bibitem{Rappoport04} T.G. Rappoport, {\it et al.},
Phys. Rev. B {\bf 69}, 125213 (2004).

\bibitem{Liu06} X. Liu and J.K. Furdyna,
J. Phys.: Condens. Matter {\bf 18}, R245 (2006).

\bibitem{Comment1} Providid that the rest of experimental conditions are identical.

\bibitem{Laletsin04} U. Laletsin {\it et al.},
Appl. Phys. A {\bf 78}, 33 (2004).

\bibitem{Comment2} Assuming an effective demagnetization factor $N=1/3$ for a randomly shaped domain.

\bibitem{Belesi12} M. Belesi {\it at al.}, 	arXiv:1204.3783v1 (2012).



\end{thebibliography}
\end{document}